# Effect of molecular structure and oxidation potential on the device performance of single-carrier organic diodes

JianChang Li, Kye-Young Kim, Silas C. Blackstock, and Greg J. Szulczewski

The Center for Materials for Information Technology and Department of Chemistry
The University of Alabama
Tuscaloosa, AL 35487

The fabrication of single-carrier organic diodes from a series of sixteen molecular materials is reported. We experimentally demonstrate how the molecular structure affects the film morphology, and how the film morphology influences the diode performance. The compounds, with moderate molecular size and dendritic structures, are shown to be more favorable for good device performance than those small molecules with symmetry structures. The device turn-on voltage is found to be strongly dependent on the molecular first oxidation potentials. Independent of different anode materials, no obvious interfacial charge/dipole effects are observed in the devices. Our results may suggest that single-carrier organic diode might offer a simple way for screeing appropriate molecular materials preferable for practical multilayer devices.

In the past decade, there have been considerable interest in using molecular materials to the fields of nanoelectronics,[1,2] optoelectronics,[3] and information storage.[4,5] To date, a great progress has been made in developing organic light-emitting diodes (OLEDs).[6-8] Most studies in this field are focused on multilayer device structures, where the device parameters are very complicated. One has to consider a lot of factors, such as the energy barriers at the metal contacts, the carrier mobility, and the thickness of each layer. So it would be difficult to study the effect of each layer on the device performance. To simplify this problem, it might be helpful to study single carrier organic diodes with only one active layer. For example, there are two papers on such single carrier polymer-based organic diodes have been reported.[9,10]

One advantage of molecular materials is their ability to be systematically tailored through chemical synthesis. This offers a good chance to study the correlation between the molecular properties and the device performance. For instance, Adachi et al. reported that the lowest drive voltage was observed in OLEDs made from hole transport materials with the lowest ionization potential.[11] In contrast, O'Brien et al. found no such relationship.[12] These conflicting reports may result from the cleaning process of ITO (indium-tin oxide) coated glass and the deposition of the top cathode.[13,14] On the other hand, though a lot of organic materials have been studied, very little information is available on how the device performance is affected by the molecular structures.

Here, we report the fabrication of single layer organic diodes from sixteen molecular materials deposited on fresh metal films. To eliminate the top electrode deposition, liquid GaIn is used as the cathode. The high work function of GaIn (~ 4.2 eV) makes holes dominate the current flow in our devices.[9] The device performance is carefully correlated with the molecular properties.

Figure 1 shows the molecular structures of the sixteen compounds used. **1** and **10–15** are designed in our lab, the synthesis of **11**, **12**, and **14** have been previously reported.[15-18] **8** was purchased from Aldrich Chemical Co. All the others were synthesized based on literature.[11] Based on the molecular structure, the materials could be intentionally divided into three groups (**1–3**, **4–9**, and **10–16**, respectively). Note that the moderate size and dendritic structures are the common characters for the molecules of the third group. As discussed later, this classification would be helpful in understanding the correspondence between the molecular structure and the device performance.

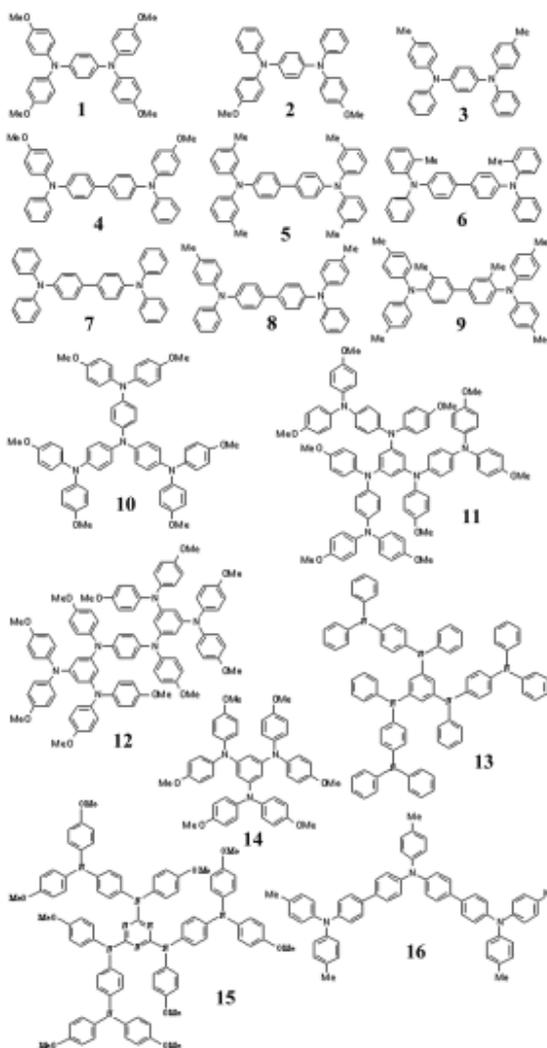

**Fig. 1 Molecular structures of the materials used.**

The depositions were conducted in vacuum chamber at a pressure of $2 \times 10^{-7}$ Torr.[19] First, silver film with thickness 100 nm was vapor deposited onto clean glass substrates at 1 Å/s and was used as anode. Without breaking vacuum, organic film was subsequently deposited onto the Ag at 0.5 Å/s. The samples were then removed and immediately measured at room temperature under ambient conditions. Liquid GaIn (75.5% Ga and 24.5% In, Alfa Aesar Co.) tip with diameter 100 μm was used as the top cathode. The current–voltage (I–V) characteristics were measured with a PAR-273 potentialstat (EG&G Instruments). The device turn-on voltage and good device rate were determined from the average of at least 50 measurements made at various positions on different samples. Low good device rate indicates high shorting rate and unstable I–V curves (and vice versa). The film morphology was imaged by atomic force microscopy at tapping mode. The molecular first oxidation potentials were measured by cyclic voltammetry.[20]

Rectification was observed in the organic diodes made from each molecular material. Figure 2 shows the representative I–V curves for six materials. Inset shows the schematic device structure and the measuring system used. Forward bias corresponds to holes emitted from the Ag anode and collected by the GaIn cathode. Negligible leakage current was measured below the turn-on voltage. Above the turn-on voltage, the current shows non-linear behavior. A preliminary analysis of these

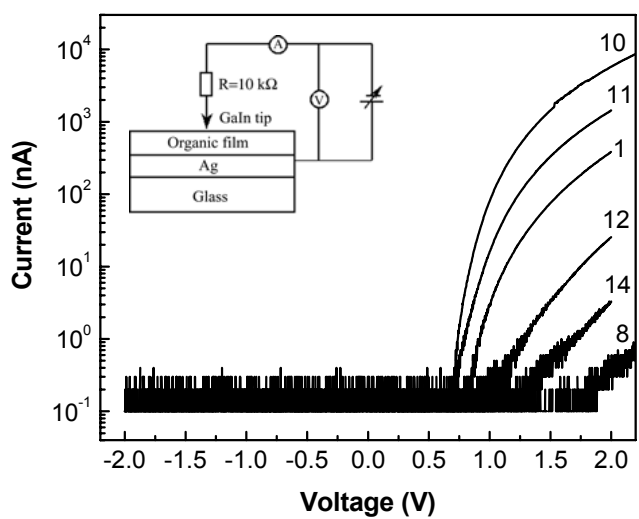

**Fig. 2 Representative I-V curves for six typical materials. Inset shows the schematic device structure and measuring system used.**

characteristics can be described by the space charge limiting conduction model.[7]

We experimentally investigated the effect of molecular structure, film morphology, and first oxidation potential on the device performance, such as good device rate, stability of the I–V curves, and turn-on voltage. The results are shown in Table I (for brevity, only **1** and **8** are given as representatives for the first two groups that show very poor device performance).

**Table I Summary of the molecular oxidation potential, film morphology, and performance of GaIn/ organic film (50nm) /Ag devices.**

| Molecule | First oxidation potential[a] (V vs. SCE) | Turn-on voltage (V) | Mean roughness[b] (nm) | Good device rate (%) |
|---|---|---|---|---|
| 1  | 0.46 | 0.79±0.09 | 75  | 10 |
| 8  | 0.85 | 1.80±0.20 | 1.8 | 25 |
| 10 | 0.32 | 0.67±0.09 | 1.7 | 80 |
| 11 | 0.45 | 0.71±0.06 | 1.0 | 95 |
| 12 | 0.49 | 1.03±0.11 | 1.4 | 85 |
| 13 | 0.64 | 1.52±0.17 | 1.9 | 30 |
| 14 | 0.67 | 1.36±0.09 | 1.1 | 85 |
| 15 | 0.71 | 1.53±0.06 | 1.2 | 70 |
| 16 | 0.74 | 1.14±0.06 | 1.8 | 90 |

[a]The first oxidation potentials of **2−7** and **9** are 0.56, 0.60, 0.76, 0.80, 0.81, 0.83, and 0.9 V vs. SCE, respectively.
[b]The mean roughness of the 50 nm films for **2−7** and **9** are about 80, 35, 1.2, 2.1, 3.9, 1.1, and 1.0 nm, respectively.

The device performance is found to be clearly correlated with the molecular structures. Briefly, the small molecular materials, with symmetry structures, tend to form rough film (**1**–**3**) or smooth films either with a lot of pinholes (**5**, **6**, **9**) or with loose-packed film morphology (**4**, **7**, **8**). Consequently, their devices show very poor performance as indicated by the low good device rates. In contrast, the films deposited from the materials of the third group, with moderate molecular size and dendritic structures, usually show both smooth and close-packed characters. As a result, most of these devices exhibit very high good device rate. It suggests that the morphology effect induced by the molecular structure plays an important role in the single carrier organic diodes.[21,22]

It is interesting to compare our results with those of the multilayer devices reported.[11] For example, **16**, with high good device rate in our diodes, exhibits high durability in the multilayer devices. Contrarily, the materials (**2**, **3**, **7**, **9**), with poor good device rate in our case, show very low durability in the multilayer devices too. So our results are well consisted with that of the multilayer devices. Moreover, it may suggest that single carrier organic diodes might provide a simple way to screen molecular materials for the study of practical multilayer devices.

Several large compounds, with higher dendritic degree and bigger molecular size than that of **11**, have been intentionally synthesized and measured. No reasonable results could be obtained due to material thermal decomposition induced by the evaporating heat. Though thin films of big dendritic molecules could be spin coated from chemical solutions,[23] that is beyond the scope of this letter.

The device turn-on voltage is found to be strongly dependent on the molecular first oxidation potentials. As shown in figure 3, the turn-on voltage seems to linearly increase as the oxidation potential increases. The error bar represents the standard deviation. Similar behavior was observed in all the compounds studied here. Note that the turn-on voltage increases more than the difference in the oxidation potentials. For example, the turn-on voltage of **11** is almost 1 V lower than that of **8**, though their oxidation potential difference is only 0.4 V vs. SCE. There are two possible contributions to such difference. One is simply the effect of the molecular oxidation potentials. The other possibility is due to the interfacial charge/dipole layers formed between the organic layer and the metal electrodes.[24]

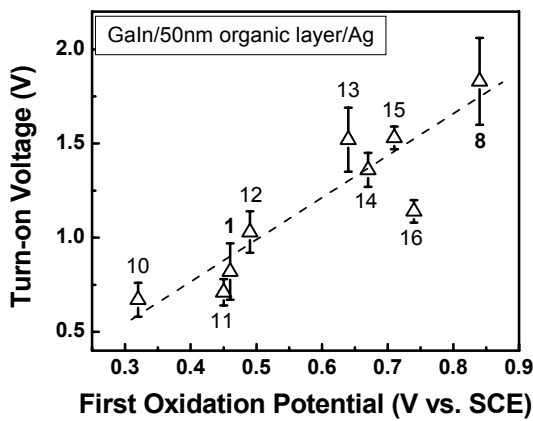

**Fig. 3 Effect of molecular oxidation potential on the device turn-on voltage. The dashed line is drawn as a guide to the eye.**

To determine which mechanism is the controlling factor, we studied the thickness dependence of the turn-on voltage for **8**, **11**, and **12**. As shown in Fig. 4, the turn-on voltage is almost linearly proportional to the film thickness. The linear extrapolation of the data to zero thickness yields an ordinate intercept of 0.6 eV for each materials studied. This value is well consistent with the work function difference between the Ag (~ 4.7 eV) anode and GaIn (~ 4.2 eV) cathode, i.e., the built-in potential.[7] Therefore, the possibility of interfacial charge/dipole effects is excluded. Otherwise, if there exist interfacial effects, the ordinate intercept should include that effect and deviate from the theoretical built-in potential.[25]

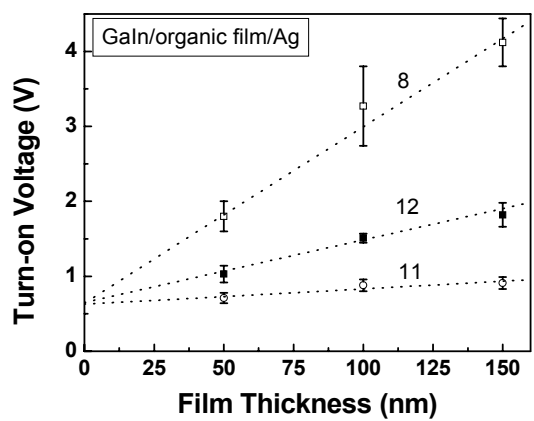

**Fig. 4 Thickness dependence of the device turn-on voltage for three representative molecules.**

To further understand the oxidation potential dependence of the turn-on voltage, theoretical calculations on representative **1**, **8**, and **11** were carried out. By using the semi-empirical method Austin Model 1, the band gap $E_g$ of **11** and **1** is calculated to be about 0.15 and 0.06 eV narrower than that of **8**, respectively. Based on the space charge limited conduction model, I–V curve simulation results indicated that the carrier mobility of **11** and **1** is about 80 and 45 times higher than that of **8**, respectively. It is known that the narrower $E_g$, the higher the intrinsic carrier concentrations.[26] Increasing the carrier mobility of the active layer could effectively lower the device turn-on voltage.[8] Accordingly, the turn-on voltage of **11** and **1** should be lower than that of **8**. Unexpectedly, the result is in well agreement with our experimental observations.

The influence of various anode materials on the device performance was studied. As shown in Fig. 5, we characterized the electrical response of **11**, exhibiting the best device performance in these materials, associated with Ag, Au, Cu, and ITO anodes. No apparent relationship between the anodes and the good device rates is observed. Independent of anodes, the thickness dependence of the device turn-on voltage shows linear behavior too. The ordinate intercepts of the turn-on voltage at zero film thickness just agree with the work function difference between the GaIn cathode and the corresponding anodes. These results confirm that there are no evident interfacial effects in our single-carrier organic devices.

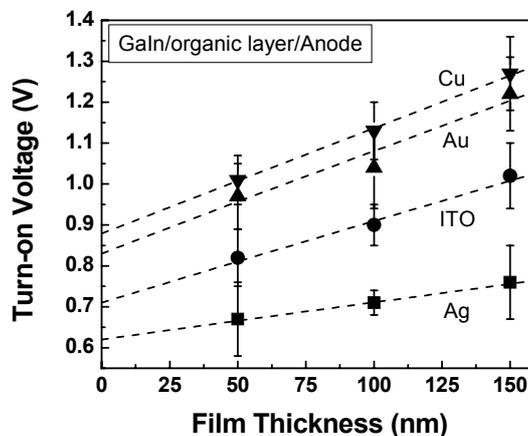

**Fig. 5** Effect of different anode metals on the device turn-on voltage for the representative molecule 11.

Based on the above experimental results, in our system, the device turn-on voltage $V_t$ can be described by

$$V_t = \Delta W + k\, d.$$

Where $\Delta W$ is the work function difference between the anode and cathode, k is a constant related to the molecular properties, d is the thickness of the organic layer. In this equation, the first term represents the built-in potential, the second part is the voltage reduction consumed in the organic layer. When both electrodes and organic layer thickness are fixed, $V_t$ could be tuned by changing the value of k (e.g. see Fig. 4). This result clearly demonstrates that how the molecular properties could play an unique role in tuning the device turn-on voltage. On the contrary, when both the electrodes and molecular material are selected, $V_t$ can only be lowered by reducing the layer thickness d (see Fig. 5). However, this will result in high leakage current and low quantum efficiency.[8]

In summary, we experimentally demonstrate that the performance of single carrier organic diodes is closely correlated with the molecular properties of the materials used, which could be attributed to the film morphology effect induced by the molecular structure. The moderate sized compounds, with dendritic structures, are shown to be favorable for good device performance. The device turn-on voltage is strongly dependent on the first oxidation potentials. Independent of anodes, no obvious interfacial charge/dipole effects were observed in the devices. We suggested that single carrier organic diode might offer a simple way for screening appropriate molecular materials preferable for ultimate practical multilayer device.

The authors would like to thank Dr. Trent Selby for synthesizing some compounds. Financial support from the National Science Foundation through the Materials and Research Science and Engineering Center (grant no. DMR-98-09423 and grant no. DMR-02-13985). One of the authors (J.C.) also thanks Prof. Z. Q. Xue for his valuable advice.

**References**

1. C. Zhou, M. R. Desphpande, M. A. Reed, L. Jones II, and J. M. Tour, Appl. Phys. Lett. **71**, 611 (1997).

2. M. A. Reed, J. Chen, A. M. Rawlett, D. W. Price, and J. M. Tour, Appl. Phys. Lett. **78**, 3735 (2001).

3. C. −Y. Liu, and A. J. Bard, Chem. Mater. **10**, 840 (1998).

4. H. J. Gao, K. Sohlberg, Z. Q. Xue, H. Y. Chen, S. M. Hou, L. P. Ma, X. W. Fang, S. J. Pang, and S. J. Pennycook, Phys. Rev. Lett. **84**, 1780 (2000).

5. J. C. Li, Z. Q. Xue, X. L. Li, W. M. Liu, S. M. Hou, Y. L. Song, L. Jiang, D. B. Zhu, X. X. Bao, and Z. F. Liu, Appl. Phys. Lett. **76**, 2532 (2000).

6. L. J. Rothberg, and A. J. Lovinger, J. Mater. Res. **11**, 3174 (1996).

7. W. Brütting, S. Berleb, and A. G. Mückl, Organic Electronics **2**, 1 (2001).

8. L. S. Hung, and M. G. Mason, Appl. Phys. Lett. **78**, 3732 (2001).

9. I. D. Paker, J. Appl. Phys. **75**, 1656 (1994).

10. P. S. Davids, I. H. Campbell, and D. L. Smith, J. Appl. Phys. **82**, 6319 (1997).

11. C. Adachi, K. Nagai, and N. Tamoto, Appl. Phys. Lett. **66**, 2679 (1995).

12. D. O'Brien, P. E. Burrows, S. R. Forrest, B. E. Koene, D. E. Loy, and M. E. Thompson, Adv. Mater. **10**, 1108 (1998).

13. J. S. Kim, M. Granstrom, R. H. Friend, N. Johansson, W. R. Salaneck, R. Daik, W. J. Feast, and F. Cacialli, J. Appl. Phys. **84**, 6859 (1998).

14. E. J. Lous, P. W. M. Blom, L. W. Molenkamp, and D. M. de Leeuw, J. Appl. Phys. **81**, 3537 (1997).

15. K. R. Stickley, T. D. Selby, and S. C. Blackstock, J. Org. Chem. **62**, 448 (1997).

16. T. D. Selby, K. −Y. Kim, and S. C. Blackstock, Chem. Mater. **14**, 1685 (2002).

17. K. R. Stickley, and S. C. Blackstock, J. Am. Chem. Soc. **116**, 11576 (1994).

18. K. R. Stickley, and S. C. Blackstock, Tetrahedron Letters **36**, 1585 (1995).

19. G. J. Szulczewski, T. D. Selby, K. −Y. Kim, J. D. Hassenzahl, and S. C. Blackstock, J. Vac. Sci. Technol. A **18**, 1875 (2000).

20. K. −Y. Kim, J. D. Hassenzahl, T. D. Selby, G. J. Szulczewski, and S. C. Blackstock, Chem. Mater. **14**, 1691 (2002).

21. M. D. Joswich, I. H. Campbell, N. N. Barashkov, and J. P. Ferraris, J. Appl. Phys. **80**, 2883 (1996).

22. Y. Shi, J. Liu, and Y. Yang, J. Appl. Phys. **87**, 4254 (2000).

23. J. M. Lupton, I. D. W. Samuel, R. Beavington, M. J. Frampton, P. L. Burn, and H. Bässler, Phys. Rev. B **63**, 155206 (2001).

24. L. Chkoda, C. Heske, M. Sokolowski, and E. Umbach, Appl. Phys. Lett. **77**, 1093 (2000).

25. K. Seki, N. Hayashi, H. Oji, E. Ito, Y. Ouchi, and H. Ishii, Thin Solid Films **393**, 298 (2001).

26. C. Kittel, *Introduction to Solid State Physics* (John Wiley & Sons, New York 1996).